\documentclass[authoryear,preprint,12pt]{elsarticle}
\usepackage{amsmath}
\usepackage{amssymb}

\usepackage{amsmath,amsthm,amssymb}
\usepackage{graphicx}
\usepackage{setspace}

\journal{Journal of Theoretical Biology}

\begin{document}

\begin{frontmatter}

\ead{zhangjiang@bnu.edu.cn}
\title{Allometry and Dissipation of Ecological Flow Networks}


\author[zhang]{Jiang Zhang}

\address[zhang]{Department of Systems Science, Beijing Normal University, Beijing}

\author[wu]{Lingfei Wu}
\address[wu]{Department of Media and Communication, City University of
Hong Kong, Hong Kong, China}

\begin{abstract}
An ecological flow network is a weighted directed graph in which
nodes are species, edges are ``who eats whom'' relationships and
weights are rates of energy or nutrients transfer between species.
Allometric scaling is a ubiquitous feature for flow systems like
river basins, vascular networks and food webs. By ``ecological
network analysis'' method, we can reveal the hidden allometry
directly on the original flow networks without cutting edges. On the
other hand, dissipation law, which is another significant scaling
relationship between the energy dissipation (respiration) and the
throughflow of any species is also discovered on the collected flow
networks. Interestingly, the exponents of allometric law ($\eta$)
and the dissipation law ($\gamma$) have a strong connection for both
empirical and simulated flow networks. The dissipation law exponent
$\gamma$ rather than the topology of the network is the most
important ingredient to the allometric exponent $\eta$. By
reinterpreting $\eta$ as the inequality of species impacts (direct
and indirect influences) to the whole network along all energy flow
pathways but not the energy transportation efficiency, we found that
as $\gamma$ increases, the relative energy loss of large nodes (with
high throughflow) increases, $\eta$ decreases, and the inequality of
the whole flow network as well as the relative importance of large
species decreases. Therefore, flow structure and thermodynamic
constraint are connected.

\end{abstract}


\end{frontmatter}


\section{Introduction}
Ecosystem is a thermodynamic system driven by energy flows which
origin from the sunlight and will be consumed by living organisms
and dissipated to the
environment\citep{odum_self-organization_1988,odum_system_1983,straskraba_ecosystems_1999}.
In this system, hundreds of species interact each other and
connected together by prey-predator interactions to form an
entangled complex network which is always called food
web\citep{pimm_food_2002,cohen_community_1990,Pascual_ecological_2005}.
In the past decades, some remarkable common patterns had been found
in binary food
webs\citep{Williams_two_2002,Sugihara_scale_1989,Bersier_scaling_1997,Krause_compartments_2003}
and also been reproduced by
models\citep{williams_simple_2000,Cattin_Phylogenetic_2004,Allesina_general_2008}
successfully. Food web, as the backbone of ecosystem, can transport
energy flow from the environment to every species as its unique
function to differentiate from other networks. In this view, food
web possesses some special features, such as low trophic
levels\citep{williams_simple_2000}, energy bottlenecks and dominator
tree\citep{Allesina_who_2004}.

Allometric scaling is a remarkable universal law found in various
flow
networks\citep{Kleiber_body_1932,West_general_1997,West_fourth_1999,Banavar_size_1999}.
The network embedded in $d$-dimensional space possesses a scaling
relation $C\propto A^{\eta}$, where $A$ is the metabolism or input
flows from the source to the network, $C$ is the total ``mass'' or
the summation of all individual flow rates in the network and
$\eta=(d+1)/d$ is the allometric exponent. River basins, mammalian
blood vessels or plant vascular systems are the flow networks in
$d=2$ and $3$ dimensional spaces
respectively\citep{Rodriguez-Iturbe_fractal_1997,Banavar_size_1999}.
\cite{Dreyer_allometric_2001} further tested this assumption in one
dimensional space by a water slot with evenly distributed sinks.
\cite{Garlaschelli_universal_2003} generalize the scaling
relationship to the spanning trees of food webs. By cutting ``weak''
links of the original food web, they calculated $A_i$ and $C_i$ for
all species on networks and found a power law relationship with
exponent around $\eta=1.13$ for almost all the food webs they
collected\citep{Garlaschelli_universal_2003,Camacho_food-web_2005,Frank_simple_2005}.

However, the mentioned food web studies always neglected the energy
flow information as the weight of links which is already available
in the empirical data\citep{brown_ecological_2003}. Studies of
ecological flow networks concerning both ``who eats whom'' binary
relationship and ``in what rate''
problem\citep{Ulanowicz_quantitative_2004} in ecology have a long
history
\citep{Finn_measures_1976,Szyrmer_total_1987,Higashi_extended_1986,Baird_seasonal_1989,higashi_network_1993,Patten_environs:_1981,Patten_environs:_1982}.
By incorporating input-output analysis and Markov chain
method\citep{Higashi_extended_1986}, ecologists have developed a
systematic approach called ``ecological network analysis'' in
ecological flow networks\citep{Ulanowicz_quantitative_2004}. They
designed a set of systematic indicators to describe the flow
structures and global state of whole
ecosystem\citep{Fath_review_1999,fath_complementarity_2001,Fath_network_1998,Hannon_structure_1973,levine_several_1980,Hannon_ecosystem_1986,Patten_energy_1985,Ulanowicz_growth_1986,Ulanowicz_ecology_1997}.

Allometric scaling as an important pattern for all flow systems
should be applied to ecological flow networks. However, due to the
limitation of the existing approach developed by Garlaschelli so
far, the extension of allometric scaling laws to any flow network is
still lack due to three major reasons: 1. The original method can be
only applied to trees so that many edges must be cut (however,
\cite{Allesina_food_2005} extended this method to directed acyclic
graphs); 2. Information on flows and weights are never considered in
previous works; 3. The ecological meaning of allometric exponent
$\eta$ should be re-considered for general flow
networks\citep{zhang_scaling_2010}.

This paper applies the energy flow analysis method to calculate the
key indicators of $A_i$ and $C_i$ for all species on flow networks
(Section \ref{sec.methods}). The allometric scaling laws for 19
empirical ecological flow networks are shown in Section
\ref{sec.allometricscaling}. Furthermore, we found that another
exponent of the scaling relationship called the dissipation law
(Section \ref{sec.dissipationlaw}) in this paper is the major
ingredient to influence the allometric exponent. We have tested this
hypothesis by a large number of numerical experiments both on
empirical and simulated flow networks (Section
\ref{sec.gammaandeta}). Finally, we re-interpret the exponent in
Section \ref{sec.discussion} as the indicator of inequality of
species impacts, i.e., the concentration degree of the species
impacts to the whole network on large species.

\section{Methods}
\label{sec.methods}
\subsection{Review of Garlaschelli's Method}
\label{sec.review}

To clarify the contribution of our method and its connection with
the existing method, we review Garlaschelli's method for a
hypothetic food web at first.

\begin{figure}
\centerline {\includegraphics{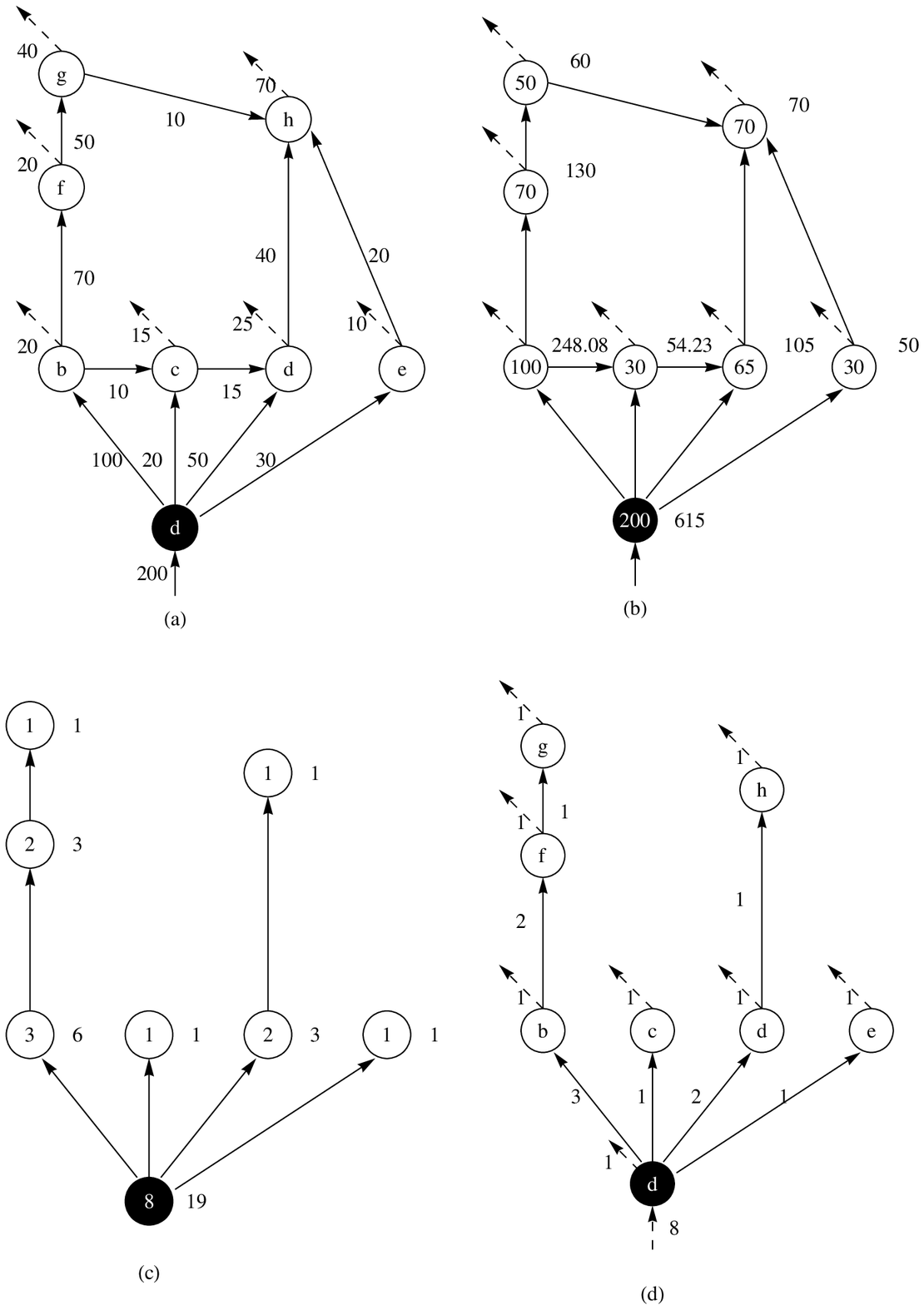}} \vskip3mm
\caption{Comparison of different methods calculating the allometric
scaling of a hypothetical ecological flow network.}{\small {\textbf
(a)} is a hypothetical network (The letter in each node is its
index). The black node is the root, the numbers besides edges stand
for flows, the dashed lines represent dissipations; {\textbf (b)}
shows the $A_i$,$C_i$ values being denoted inside and beside node
$i$ respectively by our method; {\textbf (c)} is a spanning tree of
(a), the numbers inside and beside vertex $i$ are $A_i$ and $C_i$
values calculated by Garlaschelli's method; {\textbf (c)} is the
implicated flow network of (c), the numbers beside edges are
flux}\label{fig.hfoodweb}
\end{figure}

Figure \ref{fig.hfoodweb}(a),(c) shows how Garlaschelli's approach
can be applied to a hypothetical flow network (a) to calculate $A_i$
and $C_i$ for each node. At first, a spanning tree (Figure
\ref{fig.hfoodweb} (c)) is constructed from the original network
(Figure \ref{fig.hfoodweb} (a)) by cutting edges. That way, each
sub-tree rooted from any $i$ can be viewed as a sub-system of the
spanning tree. $A_i$ is the total number of nodes involved in this
sub-tree and $C_i$ is the summation of $A_i$s for each node in this
sub-tree. Finally, the universal allometric scaling relationship of
$A_i$s and $C_i$s, with an exponent around $1.13$, was found for all
food webs according to \cite{Garlaschelli_universal_2003}.

Garlaschelli's method was inspired by \cite{Banavar_size_1999}'s
model to explain the Kleiber's law. The spanning tree is simply
Banavar's optimal transportation network. Thus, energy flows into
the whole system from the root along links of the network to all
nodes. Suppose that each node would consume 1 unit of energy in each
time step. A flux with 1 unit representing the energy consumption by
each node should then be added to the original spanning tree Figure
\ref{fig.hfoodweb}(c). In Figure \ref{fig.hfoodweb}(d), the energy
dissipation by each node is added as a dotted
line\citep{Banavar_size_1999}. As a result, $A_i$ of each node is
the throughflow of this node. $C_i$ is the total throughflows of the
sub-tree rooted from $i$. Essentially, calculation of allometric
scalings using Garlaschelli's approach is based on this weighted
flow network model\citep{zhang_scaling_2010}.

\subsection{Ecological Network Method}
We will extend Garlaschelli's method directly on the original
weighted network without cutting edges (e.g. Figure
\ref{fig.hfoodweb}). But the key question is how to calculate $A_i$
and $C_i$ for general flow network?

According to the flow network interpretation of Garlaschelli's
method in Figure \ref{fig.hfoodweb} (d), $A_i$ is the energy flux
intake by node $i$ which is equals to the total out flows from $i$
due to the flow balance condition. Thus, this concept can be
extended to any flow network by defining $A_i$ as the throughflow of
node $i$. However, defining $C_i$ is not as simple as $A_i$ because
we don't have the sub-system concepts if the considered network is
not a tree.

To understand what does $C_i$ mean in Garlaschelli's method based on
the flow network picture (Figure \ref{fig.hfoodweb} (d)), we suppose
to do the following hypothetic experiment. Assume a large number of
particles are flowing along the network (Figure \ref{fig.hfoodweb}
(d)), and all particles whenever passing by any node, say b, will be
attached a label, say ``b''. This trace marker will not be erased
forever unless the particle flows out of the network. Then, we found
$C_b$ in Figure \ref{fig.hfoodweb} (c) is just the total number of
labeled particles by ``b'' still being trapped in the whole network.
This trace marker experiments can be also applied to other nodes
independently and separately by attaching different labels so that
all $C_i$s can be calculated by counting the number of particles who
used to pass node $i$.

Actually, this understanding can be extended to any flow network no
matter if it is a tree or not\citep{zhang_scaling_2010}. Although
counting the number of labeled particles in the real network is
difficult, we can do this calculation directly by Markov chain
technique thanks to the ecological network analysis method developed
by Patten et al. As long as the flow network is in the steady state
so that all flows distributed on edges are stable, a fixed $C_i$
value can be calculated according to the flow structure.

Suppose the flux matrix of the original network is $F$, in which
each entry $f_{ij}$ stands for the flux from $i$ to $j$. Two special
nodes $0$ and $N+1$ ($N$ is the total number of species)
representing source and sink are contained in this matrix as the
first(last) in columns(rows). We define $A_i$ of $i$ as,
\begin{equation}
A_i=\sum_{j=1}^{N+1}{f_{ij}},  \forall i \in [1,N]
\end{equation}
That is the throughflow of species $i$ in ecological network
analysis\citep{Ulanowicz_ecology_1997}.

To calculate $C_i$, we should convert the original flux matrix $F$
into a Markov chain in which each element is defined as
$m_{ij}={f_{ij} /(\sum_{k=1}^{N+1}{f_{ik}})}$ for all $1\leq i\leq
N$. Notice that the Markov chain is normalized (i.e.
$\sum_{j=1}^{N+1}{m_{ij}=1}$) only if the original flux matrix is
balanced, i.e.,
\begin{equation}\label{eqn.balanced}
\sum_{i=0}^{N+1}{f_{ij}}=\sum_{j=0}^{N+1}{f_{ij}}, \forall i \in
[1,N]. \end{equation}

This requirement is always satisfied by most empirical ecological
networks. The webs not satisfying this condition will be balanced by
the method mentioned in Appendix \ref{sec.balance}. Thus, $C_i$ can
be calculated as,

\begin{equation} \label{eqn.ci}
C_i=\sum_{k=1}^N{\sum_{j=1}^N{f_{0j}\frac{{u_{ji}u_{ik}}}{u_{ii}}}},
\end{equation}
where, $u_{ij}$ is the element in matrix $U$, which is called
fundamental
matrix\citep{Fath_review_1999,Ulanowicz_quantitative_2004} and
defined as,

\begin{equation} \label{eqn.umatrix}
U=(I-M)^{-1},
\end{equation}
where, $I$ is the identity matrix. According to the ecological
network analysis method, $C_i$ is just the total number of particles
that used to pass node $i$ (see Appendix \ref{sec.cicalculation}).

Then, we can calculate $A_i,C_i$ values for all nodes in the flow
network to test the following allometric scaling law,
\begin{equation} \label{eqn.umatrix}
C_i\propto A_i^{\eta}.
\end{equation}

Where, $\eta$ is the allometric exponent that will be mainly
discussed in the following sections.

\section{Results}
\label{sec.results}
\subsection{Description of Data Set}
The 19 ecological flow networks in different habitats are
studied(Table \ref{tab.foodwebdata}). The networks are obtained from
the online database\footnotemark.
\footnotetext[1]{http://vlado.fmf.uni-lj.si/pub/networks/data/bio/foodweb/foodweb.htm}Most
of these networks in this database are from the published
papers\citep{almunia_benthic-pelagic_1999,Baird_seasonal_1989,Baird_assessment_1998,Hagy_Eutrophication_2002,Ulanowicz_growth_1986,Monaco_comparative_1997,Christian_organizing_1999}.
In Table \ref{tab.foodwebdata}, we list the name, the number of
nodes ($N$) and the number of edges ($E$) of these networks. In
which, the nodes are living species and also non-living compartments
(e.g., DOC, POC), the weighted links are energy flows whose values
vary in a large range because the units and time scales of the
measurements are very different. The source node $0$ and sink node
$N+1$ are the ``input'' node and the combination of ``respiration''
and ``output'' nodes in the original data respectively. The
dissipative flux of each node $i$ is just the flow from $i$ to $N+1$
which can be read from the data directly. Most ecological flow
networks are balanced already (condition Equation
\ref{eqn.balanced}), few imbalanced networks are balanced by the
approach mentioned in Appendix \ref{sec.balance}.

\begin{table}
\centering
 \caption{Empirical Ecological Flow Networks and Scaling
 Exponents}{\small{$N$ and $E$ are the total numbers of nodes and edges respectively, all networks are sorted in decreasing order of $N$} }
 \label{tab.foodwebdata}
\begin{tabular}{cccccccc}
\hline Food web & Abbre.  & $N$ & $E$ & $\eta$ & $R^2_{\eta}$ & $\gamma$ & $R^2_{\gamma}$\\
\hline
\small{Florida Bay, Dry Season}   &Baydry&126&2102&1.010&0.995&0.915&0.949\\
\small{Florida Bay, Wet Season}   &Baywet&126&2071&1.020&0.994&0.917&0.953\\
\small{Mangrove Estuary,}\\\small{Dry Season}   &Mangdry&95&1462&1.010&0.997&0.978&0.983\\
\small{Everglades Graminoids,}\\ \small{Wet Season}   &Gramdry&67&863&1.030&0.999&0.973&0.997\\
\small{Everglades Graminoids,}\\ \small{Wet Season}   &Gramwet&67&863&1.020&0.999&0.977&0.998\\
\small{Cypress,Dry Season} & CypDry  &69&639&0.998&0.996&0.957&0.949\\
\small{Cypress,Dry Season} & CypDry  & 70 & 554&0.998&0.996&0.967&0.949\\
\small{Cypress,Wet Season} & CypWet  &69&630&0.997&0.997&0.965&0.988\\
\small{Mondego Estuary}\\\small{-Zostrea site}& Mondego  &44&401&1.010&0.999&0.979&0.997\\
\small{St. Marks River (Florida)} & StMarks &52&349&1.020&0.980&0.985&0.950\\
\small{Lake Michigan} & Michigan  &37&210&1.010&0.999&0.995&0.999\\
\small{Narragansett Bay} & Narragan  &33&194&1.010&0.991&0.813&0.942\\
\small{Upper Chesapeake}\\\small{ Bay in Summer} & ChesUp &35&203&1.050&0.997&0.952&0.991\\
\small{Middle Chesapeake}\\\small{ Bay in Summer} & ChesMiddle &35&195&1.040&0.996&0.851&0.761\\
\small{Chesapeake Bay}\\ \small{Mesohaline Net} & Chesapeake &37&160&0.994&0.997&0.985&0.985\\
\small{Lower Chesapeake}\\\small{ Bay in Summer} & ChesLower  &35&163&1.050&0.997&0.926&0.971\\
\small{Crystal River Creek}\\\small{ (Control)}   &CrystalC&22&107&1.040&0.997&0.959&0.995\\
\small{Crystal River Creek}\\ \small{(Delta Temp)}   &CrystalD&22&83&1.040&0.998&0.963&0.996\\
\small{Charca de}\\ Maspalomas & Maspalomas  &22&82&0.956&0.966&1.150&0.737\\
\small{Rhode River Watershed}\\ \small{- Water Budget}  &Rhode&18&54&0.828&0.866&1.200&0.963\\
 \hline
\end{tabular}
\end{table}

\subsection{Allometric Scaling Law}
\label{sec.allometricscaling}

\begin{figure}
\centerline {\includegraphics{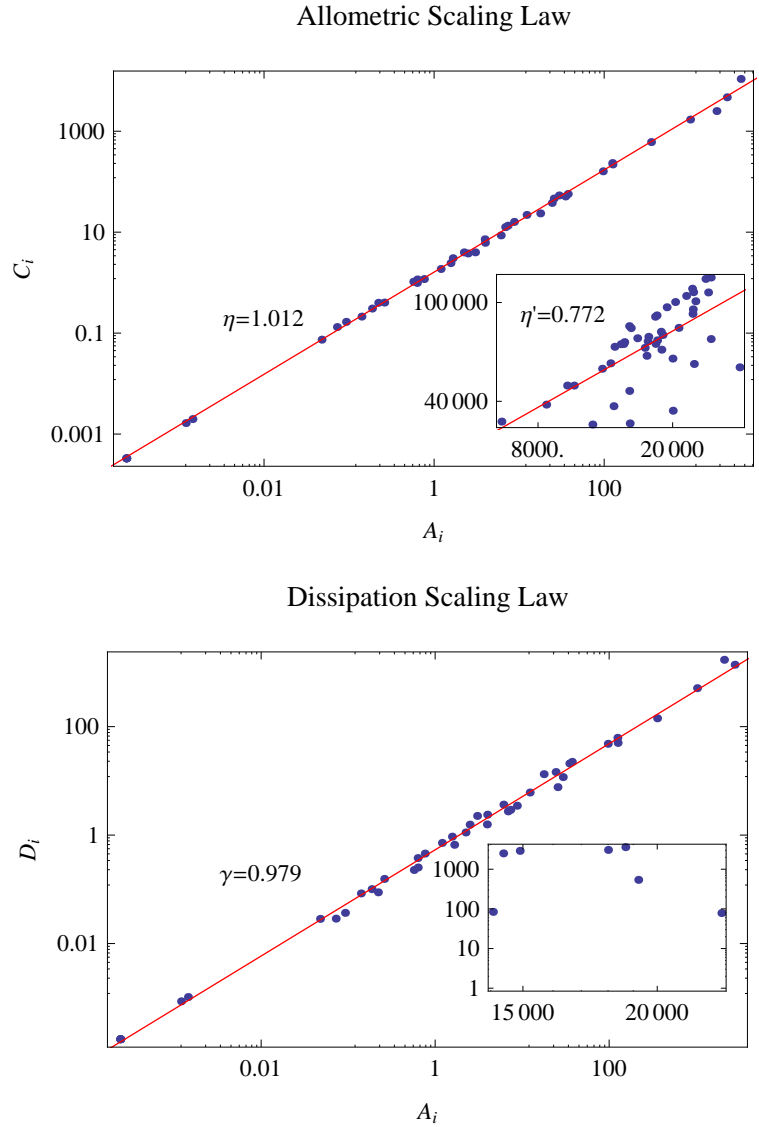}} \vskip3mm
\caption{Allometric scaling law and dissipation law for Mondego
ecological network and the null model (insets)}{\small{There are
only few points in the inset of the lower figure because many nodes
in the null model are not balanced and their dissipations are set to
zeros}}\label{fig.scalinglaw}
\end{figure}

We found all the ecological flow networks possess the allometric
scaling pattern significantly as an example shown in Figure
\ref{fig.scalinglaw}. Their allometric exponents $\eta$s with the
values of $R_{\eta}^2$ are listed in Table \ref{tab.foodwebdata}. We
see that the values of $R_{\eta}^2$ are larger than $0.9$ for all
food webs except CrystalC, CrystalD and Rhode whose scales are very
small ($N<23$). All of exponents $\eta$ fall into the interval
$[0.83,1.05]$, most of them are larger than $1$ a little.

To test if the allometric scaling pattern is significant compared to
random flow networks, we built a null model in which the numbers of
nodes and edges are kept, all links are re-connected randomly and
all flows on edges are also randomly assigned on the interval
$(0,f_m]$ evenly, where $f_m$ is the maximum flux of the original
network. From the inset of first row in Figure \ref{fig.scalinglaw},
we see the null model network doesn't show the allometric scaling
law. We also try to compare the empirical flow networks with other
null models. Some of them are obtained by keeping the topology
unchanged but randomly assigning the weights, some of them are
obtained by just shuffling the weights among edges, all the details
of these null models are presented in the Appendix
\ref{sec.nullmodels}.

Among these null models, we found the ones keeping weights
information have similar allometric scaling exponent as the the
empirical food webs. As a result, we know that it is the flow
distribution but not the topological structure of the network that
is the key ingredient to the allometric scaling exponents. However,
to study all possible flow distributions which can affect the final
allometric law is impossible because there are hundreds of flows
which can be adjusted for an empirical food web. What is the most
important aspect? We found the dissipative flux is the key to the
allometric exponent.

\subsection{Dissipation Law}
\label{sec.dissipationlaw}

In ecology, dissipation of a species has different forms such as
respiration, excretion, egestion, natural and predatory mortality
and so forth\citep{straskraba_ecosystems_1999}. In our data, the
dissipative flow is mainly respiration. We can understand the
dissipation of a species $D_i$ in an ecological flow network as the
flows out of the network, i.e., $D_i=f_{i,N+1}$. It is
comprehensible that this output flow increases with the total
throughflow of the focus species. But it is not obvious that for
most collected ecological flow networks the growth of dissipation
along different species is slower than the growth of throughflow. A
sub-linear relationship between dissipation and throughflow of each
species is hold per se,
\begin{equation} \label{eqn.dissipationlaw}
D_i\propto A_i^{\gamma},
\end{equation}
where, the exponent $\gamma$ in Equation \ref{eqn.dissipationlaw} is
called dissipation law exponent which can be estimated from the
data. Its values for different networks are also listed in Table
\ref{tab.foodwebdata}. We see all the $\gamma$ values are smaller
than 1 except Maspalomas and Rhode networks. The lower plot in
Figure \ref{fig.scalinglaw} shows the dissipation scaling law for
Mondego flow network as an example.

\subsection{Relationship Between $\gamma$ and $\eta$}
\label{sec.gammaandeta}

\begin{figure}
\centerline {\includegraphics{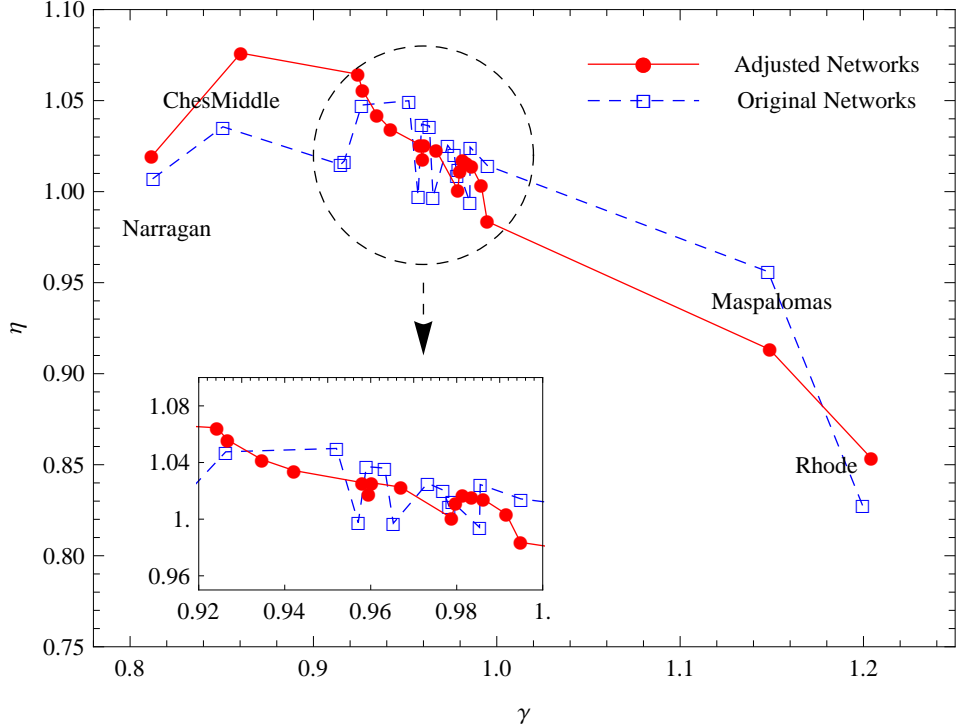}} \vskip3mm
\caption{The relationship between $\gamma$ and $\eta$ in original
and adjusted ecological flow networks}{\small{The blue dashed line
is the $\gamma-\eta$ curve for the original flow networks, and the
solid red one is for the adjusted networks based on the same network
structure and the dissipation law according to flow adjusted
algorithm (see main text)}}\label{fig.gammaandeta}
\end{figure}

Because both $\gamma$ and $\eta$ are indicators for the whole flow
network, to see how these two numbers correlate each other, we can
simply plot different $\eta$s against $\gamma$s across all the
collected empirical flow networks(see the blue dotted curve in
Figure \ref{fig.gammaandeta}). Although a general trend that $\eta$
decreases with $\gamma$ can be observed, it is not significant due
to three major reasons. First. The number of sample points is too
small to show clear relationship. Second, most of data points are
concentrated in the circled area since all $\gamma$s and $\eta$s are
of similar values for all networks. Finally, the original exponents
are noisy so that the $\eta$ values fluctuate in the circled area.

To understand how $\eta$ depends on $\gamma$ for a given flow
network structure as well as to avoid the problems mentioned in the
previous paragraph, we invent a specific technique called ``Flow
Adjusting Algorithm (FAA)''. This algorithm enables us to perturb
the given flow network structure as little as possible and
simultaneously observe how $\eta$ changes by tuning $\gamma$.
Concretely, for a given original flow network $F$, we keep the
network topology, the relative importance of each influx unchanged
(That is, $f_{ji}/\sum_{j}{f_{ji}}=f'_{ji}/\sum_{j}{f'_{ji}}$),
meanwhile adjust the flow distributions on each edge to obtain a new
flow network $F'$ such that:

(1) The given dissipation law, i.e., $D'_i\propto
{A'_{i}}^{\gamma}$, is hold for every node $i$ in $F'$, where
$\gamma$ is a given exponent which can be tuned. We will study how
$\gamma$ impacts the allometric scaling law;

(2) The flux balance condition, i.e.,
$\sum_{j}f'_{ij}=\sum_{j}f'_{ji}$ must be kept for each node $i$;

In this way, we can perturb the flow structure of the original
network to obtain the expected dissipation law, and to observe how
exponent $\gamma$ affects exponent $\eta$. The details of the ``flow
adjusting algorithm'' will be introduced in Appendix
\ref{sec.flowadjustingalgorithm}. Figure \ref{fig.flowadjusting}
shows the dependence of $\eta$ and $\gamma$ on the perturbed
networks by MAA based on the original Mondego flow network,
randomized Mondego flow network (based on Mondego's topology but
assign flows randomly) and simulated networks by Niche
model\citep{williams_simple_2000}. We observe the allometric scaling
exponent decreases with the dissipation law exponent in a similar
manner no matter the original network structures are. However, the
concrete shapes of the curves between $\eta$ and $\gamma$ change
with the network structures. For networks generated by Niche model,
$\eta$ decreases with $\gamma$ in a slower speed when the
connectances are higher (blue triangles versus purple triangles and
yellow diamonds versus green diamonds). As a result, the dissipation
law exponent but not the structure is the major feature to affect
the allometric exponent. But we cannot conclude the topological
structure has no influence on the allometric exponents, we will
discuss this problem further in Appendix
\ref{sec.simulatednetworks}.

\begin{figure}
\centerline {\includegraphics{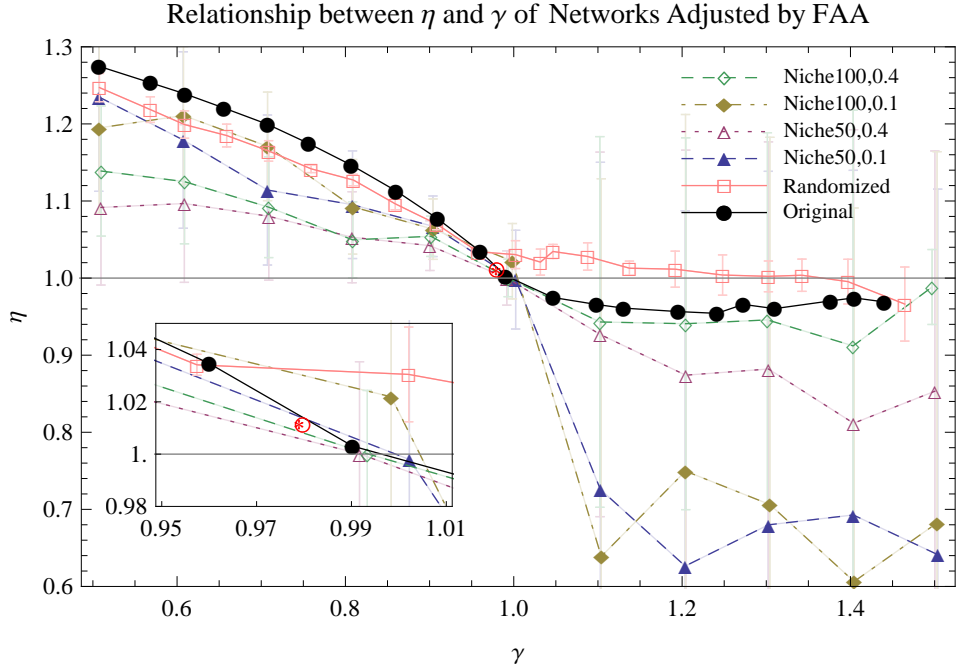}} \vskip3mm
\caption{The relationship between $\gamma$ and $\eta$ by flow
adjusting algorithm applying on Mondego ecological network,
randomized Mondego flow network, and generated networks by Niche
Model.}{\small{The black and red curves are the original and
randomized flow networks of Mondego respectively. The randomized
flow network keeps the topological structure of Mondego network but
assign flux randomly. Other dashed curves are for networks generated
by Niche model with different number of nodes (100 or 50) and
connectances (0.4 or 0.1), as well as the random assigned
flows\citep{williams_simple_2000}. MAA is applied on all these flow
networks to obtain the relationships between $\eta$ and $\gamma$.
All the exponents are the average results of 50 random experiments.
The red star shows the original position of dissipation law exponent
(0.980) and allometric exponent (1.012) of Mondego network. While
the red circle corresponds to the combination of the original
dissipation law exponent (0.980) and the adjusted allometric
exponent (1.011) by FAA on Mondego network.}
}\label{fig.flowadjusting}
\end{figure}

In Figure \ref{fig.flowadjusting}, the red circle in the middle of
the black curve stands for the combination of the dissipation law
exponent $\gamma$ of the original Mondego network $F$ and the
allometric exponent $\eta'$ of the adjusted flux matrix $F'$ by the
flow adjusting algorithm. While, the red star corresponds to the
original exponents of Mondego network both for $\gamma$ and $\eta$.
We see these two points almost overlap together which means the
original Mondego food web satisfies two conditions:

(1) dissipation law is significant for the original $\gamma$;

(2) the balance Equation \ref{eqn.balanced} is obeyed exactly.

However, for other empirical networks, the perturbed result of
allometric exponent is not identical to the original one exactly on
the given original dissipation law exponent because either the flux
balance condition is violated or the dissipation scaling law is not
significant (see Appendix \ref{sec.empiricalnetwork}).

By adjusting the flows on all empirical flow networks with the fixed
original dissipation law exponent $\gamma$, we can eliminate noise
in raw data because we conform the network to be flow balanced and
satisfy the given dissipation law. The red solid curve in Figure
\ref{fig.gammaandeta} shows the clear dependence of $\gamma$ and
$\eta$ of all the empirical flow networks. We see the decreasing
trend of the blue dashed line standing for the original exponents is
not as obvious as the red one representing the adjusted results.
Therefore, the two conditions listed before cannot be satisfied by
the some original ecological networks perfectly.

\section{Discussion}
\label{sec.discussion}
\subsection{Transportation Efficiency or Inequality of Species Impact?}
Previous studies explained the allometric scaling exponent $\eta$ as
the transportation efficiency of the network and is bound in between
1 (a star network, the most efficient tree) and 2 (a chain, the most
inefficient tree).

However, in our study, the allometric scaling exponent is not bound
in the interval $[1,2]$, therefore, we should give a new explanation
for the exponent $\eta$. The key problem is to understand the
indicator $C_i$.

In Banavar's model and Garlaschelli's method, they understand $C_i$
as a cost of energy transportation for the sub-tree rooted from $i$.
In this way, the energy flows on the redundant links (loops or
cross-leveled links) except the ones in the spanning tree are
wasted. Nevertheless this interpretation can hardly generalize to
flow networks because (1) the wasted energy in the weighted networks
can be measured as dissipation of each node but not the weight of
edges; (2) all energy links should be considered because they all
contribute to the whole flow distribution.

According to the particle coloring experiment mentioned in Section
\ref{sec.methods}, we can understand $C_i$ as the total impact of
$i$ to the whole network along all flow
pathways\citep{Vitali_network_2011} because it is the total number
of particles who used to pass $i$ at least one time. Thus, as $C_i$
increases, more other nodes will be affected by the particles used
to pass $i$, the direct and indirect influences of $i$ will
increases. This understanding toward $C_i$ can be extended to any
flow network.

As the species climbs up along the energy throughflow gradient
$A_i$, its total impact $C_i$ also increases with a relative speed
$\eta$ according to the allometric scaling law $C_i\propto
A_i^{\eta}$. Therefore, the important nodes (with larger $A_i$) may
have much greater power (total impact $C_i$) in the networks with
larger exponent $\eta$ than those networks with smaller exponents.
For example, we have two networks with 4 nodes. They have the same
throughflow distributions, e.g., $A_i=\{1,2,3,4\}$, but different
exponents $\eta_1=1$ and $\eta_2=2$, therefore different $C_i$
distributions $C_i^{(1)}=\{1,2,3,4\}, C_i^{(2)}=\{1,4,9,16\}$. The
most important node (with largest $A_4=4$) in the second network may
have much greater impact ($C_4^{(1)}=16$) to the whole network than
the first one ($C_4^{(1)}=4$). Hence, the inequality of the species
impact of the second network is larger than the first one. In short,
the allometric exponent $\eta$ measures the inequality of species
impact.

This new interpretation is compatible with the previous one. For
Garlaschlli's spanning trees, the most inequable tree with a given
root is a chain but the most equable one is a star.

\subsection{Dissipation and Inequality}
According to the new interpretation of allometric exponent and the
discovery of the relationship between dissipation and allometry, we
can obtain a whole picture: the networks may become more equable by
dissipating more energy on larger nodes since the impacts of high
flux nodes are weakened. In the networks with $\gamma>1$, the
dissipation flow per throughflow of each node increases, the energy
invested to the whole network decreases with the size of node, the
network is more decentralized. On the other hand, if $\gamma$ is
smaller than 1, the dissipating flux scales to the throughflow with
a smaller relative speed. So the large nodes may input more energy
on the whole network to obtain much more powerful impact on other
nodes, the networks are more centralized.

However, an interesting unexplained fact is the allometric exponents
of empirical ecological networks are all close to 1. They are
neither inequable nor equable. We guess this exponent should be an
optimal result by some unknown factors. This will be left for future
studies.

In summary, this paper generalize the universal allometric law to
the ecological flow networks and discover that the major factor to
the allometric exponent is $\gamma$, the dissipation law exponent.
By reinterpreting allometric exponent as the inequality of species
impacts, we build a connection between network structure and the
thermodynamic constraint. This connection is very important
deserving more attention.

\bibliographystyle{elsarticle-harv}
\bibliography{ecology}

\paragraph{{\large{Appendix}}}
\appendix
\section{Balancing A Flow Network}
\label{sec.balance} For most of empirical ecological flow networks,
the flux matrix $F$ is balanced which means
$\sum_{j=0}^N{f_{ji}}=\sum_{j=1}^{N+1}{f_{ij}}$ holds for each $i\in
[1,N]$. However, some empirical networks and most artificial
networks (e.g. random network) are imbalanced. Therefore, we should
balance the given network $F$ artificially so that the ecological
network analysis methods can be applied.

Suppose $\sum_{j=0}^{N}{f_{ji}}\neq \sum_{j=1}^{N+1}{f_{ij}}$ for
node $i$. We can add an edge with the flux $|w|$,
$w=\sum_{j=0}^{N}{f_{ji}}-\sum_{j=1}^{N+1}{f_{ij}}$ to connect node
$i$ to $N+1$ or $0$. If $w>0$, the direction of this artificial edge
is from $i$ to $N+1$. If $w<0$, the direction is from $0$ to $i$. We
can do this process for all nodes except $0$ and $N+1$ to balance
the whole network.

\section{Explanation on $C_i$ Calculation by an Example Network}
\label{sec.cicalculation} 
\begin{figure}
\centerline {\includegraphics{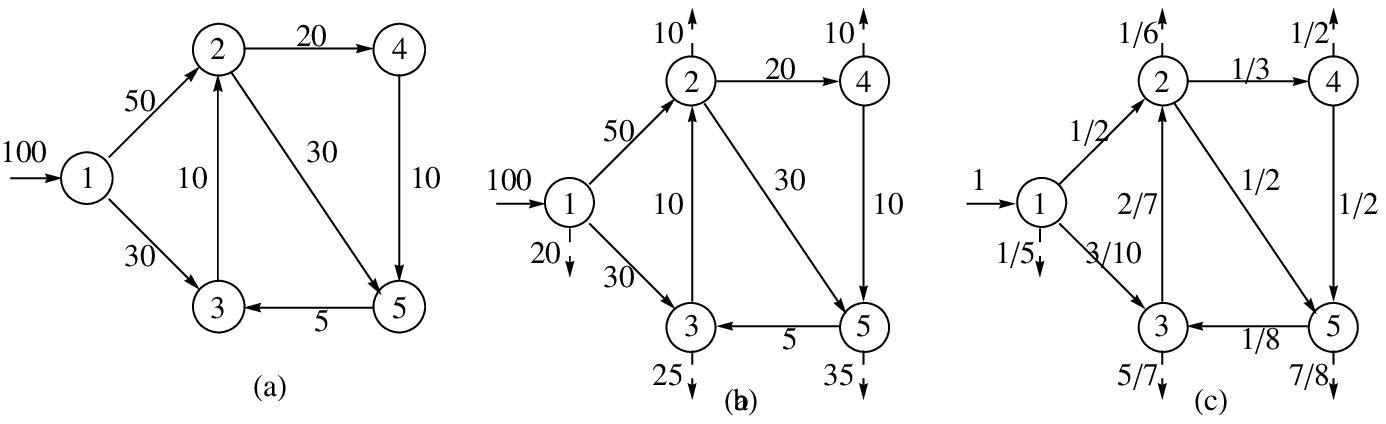}} \vskip3mm \caption{An
example network showing the calculations of $A_i$ and
$C_i$}{\small{(a). The original flow network which is imbalanced;
(b). The balanced network; (c). The corresponding Markov chain}
}\label{fig.example}
\end{figure}

In this Section, we will explain why the
number of particles labeled by node $i$ can be calculated as
Equation \ref{eqn.ci} through an example network (see Figure
\ref{fig.example}).

Notice that the original network is imbalanced, we should balance it
at first by the approach mentioned in the last section. Then, the
balanced network can be converted to a Markov chain $M$ simply
through being normalized by the output flow of each node as shown in
Figure \ref{fig.example} (c). We know any element $m_{ij}$ in Markov
matrix $M$ stands for the transfer probability of a particle from
$i$ to $j$ given that the particle locates $i$ already and the whole
network is in the steady state. And any element in matrix $M^t$ is
the probability of a particle locating on $i$ at first and transfer
to $j$ along all possible pathes after $t$ steps. All these
information is aggregated in the fundamental matrix $U$ because,

\begin{equation}
\label{eqn.fundamental}
U=I+M+M^2+\cdot\cdot\cdot+M^{\infty}=(I-M)^{-1}.
\end{equation}

We should be careful to give an intuitive explanation on $U$ because
its element $u_{ij}$ does not stand for probability anymore although
the element in each term $M^t$ is the transitional probability. We
can write down the fundamental matrix for our example network (by
ignoring the elements for source $0$ and sink $N+1$),

\begin{equation}
\label{eqn.exampleu}U = I+M+M^2+ \cdots=\left(
\begin{array}{ccccc}
 1 & 3/ 5 & 7/ 20 & 1/ 5 & 2/ 5\\
 0 & 42/ 41 & 7/82 & 14/41 & 28/41\\
 0 & 12/41 & 42/41 & 4/41 & 8/41\\
 0 & 3/164 & 21/328 & 165/164 & 21/41\\
 0 & 3/82 & 21/164 & 1/82 & 42/41
\end{array}
\right)
\end{equation}

Notice that the elements on the diagonal are larger than 1, so they
cannot be interpreted as probability simply.

Next, we will calculate the first passage flow $G_{i}$ to any node
$i$. Here, $G_{i}$ is defined as the number of particles passing $i$
in the first time. If all the particles passed $i$ are labeled, then
$G_{i}$ is the number of particles passing $i$ and unlabeled by it
at each time. This quantity can be calculated as,

\begin{equation}
\label{eqn.firstpassage} G_{i}=\sum_{j=1}^{N}\frac{f_{0j}
u_{ji}}{u_{ii}}.
\end{equation}

For example, the first passage flow of node 4($G_{4}$) can be
calculated as,

\begin{equation}
\label{eqn.firstpassage1}
G_{4}=\sum_{j=1}^{N}\frac{f_{0j}u_{j4}}{u_{4,4}}=f_{0,1}\frac{u_{1,4}}{u_{4,4}}=100\times
\frac{1/5}{165/164}={656}/{33}
\end{equation}

Actually, each term of Equation \ref{eqn.firstpassage} is the first
passage flow from node $j$ to $i$, that is the number of particles
that have visited $j$ (in whatever time) and finally arrive at $i$
in the first time along all possible flow pathways. By dividing by
the term $u_{ii}$ to derive first passage flow one avoids duplicate
counting the particles who have visited
$i$\citep{higashi_network_1993}.

Because at each time, there are totally $G_{i}$ unlabeled particles
will pass $i$ and be unlabeled by ``b'', thereafter, $G_{i}$ new
``b'' particles will be injected to the system and will flow to
other nodes along pathways. Then, at given time step, there are
totally $G_i$ labeled particles just injected $i$, and
$\sum_{k=1}^{k=N}{G_{i} M_{\{ik\}}}$ labeled particles injected one
time step ago, and $\sum_{k=1}^{k=N}{G_{i}M^2_{\{ik\}}}$ labeled
particles injected two time steps
ago,...,$\sum_{k=1}^{k=N}{G_{i}M^t_{\{ik\}}}$ labeled particles
injected $t$ time steps ago, and so forth.

Hence, the total number of labeled particles that flowing in the
whole network at each time, being defined as $C_i$, is just the
summation of the labeled particles being attached 1 time step ago, 2
time steps ago,..., and so on. Therefore, we can calculate $C_i$ as:
\begin{equation}
\label{eqn.ciunderstanding} \begin{aligned}C_i&=\sum_{k=1}^{N}
{G_i(I+M+M^2+\cdot\cdot\cdot+M^{\infty})_{\{ik\}}}\\&=\sum_{k=1}^{N}{G_i
U_{\{ik\}}}\\&=\sum_{k=1}^{N}{(\sum_{j=1}^{N}{f_{0j}u_{ji}/u_{ii}})u_{ik}}
\end{aligned}
\end{equation}

For example, $C_2$ of node 2 in the example network is calculated
as:
\begin{equation}
\label{eqn.exampleC} C_2=G_2 \sum_{k=1}^5{ u_{2k}}=((100\times 3/5 +
0)/ (42/41))\sum_{k=1}^5{u_{2k}} =125
\end{equation}

\section{Null Models}
\label{sec.nullmodels} To test if the allometric scaling law is a
significant pattern in empirical ecological network, we designed
four kinds of null models based on empirical networks.

Null Model 1 (NM1): We only keep the total number of nodes and edges
as the original empirical flow network, and build random
connections, assign random weights for each edge. The weights are
evenly distributed on the interval $(0,f_m]$, where $f_m$ is the
maximum flux in the original network. In this way, the topology and
flow distribution are destroyed.

Null Model 2 (NM2): The connections are kept, the weights are
randomly assigned for each edge. In this model, weights are also
randomly sampled from the interval $(0,f_m]$. In this way, only the
flow distribution is destroyed.

Null Model 3 (NM3): Keep the connections, shuffle the weights on
edges. That is, we keep the topology and weights distribution but
permute these weights on edges. In this way, the flow distribution
is not changed but the correlations between flows are destroyed.

Null Model 4 (NM4): Keep the weights, the number of edges, but
randomly assign the weighted connections between any pair of nodes.
In this way, the flow distribution is kept, but their correlations
and the network topology are destroyed.

For each original ecological flow network, we built four null
models. The flux matrix $F$ of the null models is imbalanced
normally, then we should balance it by the approach mentioned in
Section \ref{sec.balance}, after that we calculate their $A_i$ and
$C_i$ for each node and derive the allometric scaling law pattern.

\begin{figure}
\centerline {\includegraphics{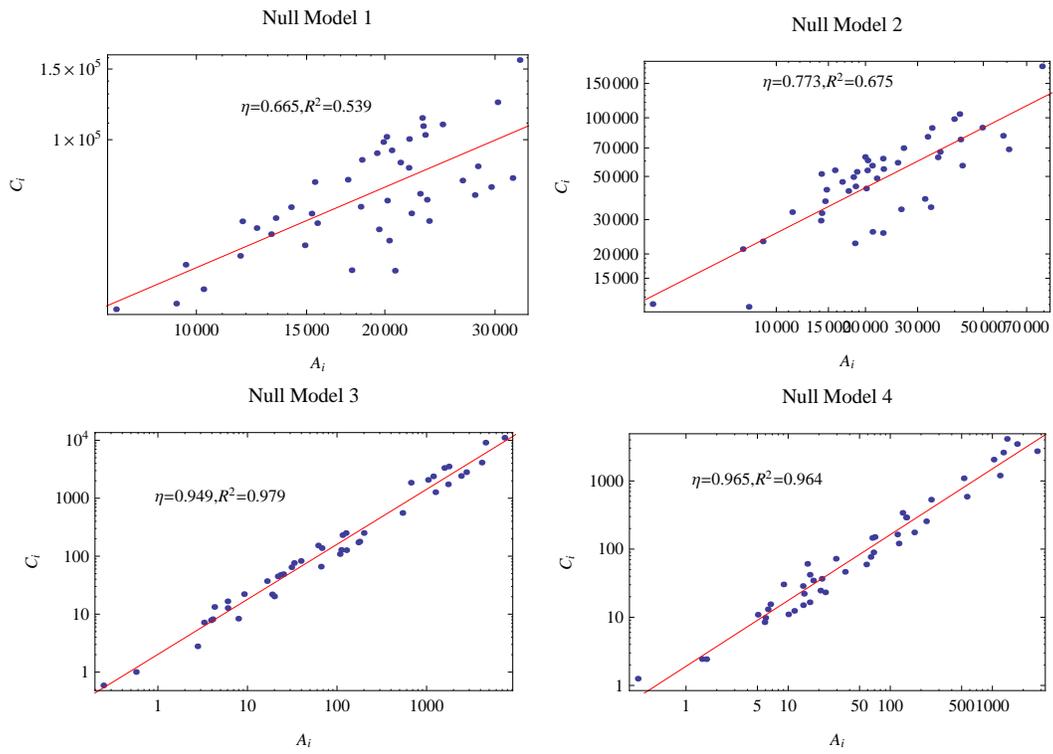}} \vskip3mm
\caption{Allometric Scaling Patterns for Null Models based on
Mondego Network}\label{fig.nullmodels}
\end{figure}

Figure \ref{fig.nullmodels} shows the allometric scaling
relationships for $A_i$s and $C_i$s for null models of Mondego
network. From this figure, we know NM3 and NM4 have more similar
pattern as the original networks than NM1 and NM2, which means the
flow distribution is more important than topology for allometric
scaling. Although NM3 and NM4 have significant scaling pattern,
their exponents $\eta$s are smaller than the one of the original
Mondego network. Therefore, NM3 and NM4 cannot reproduce the main
characters of the original network.

\begin{figure}
\centerline {\includegraphics{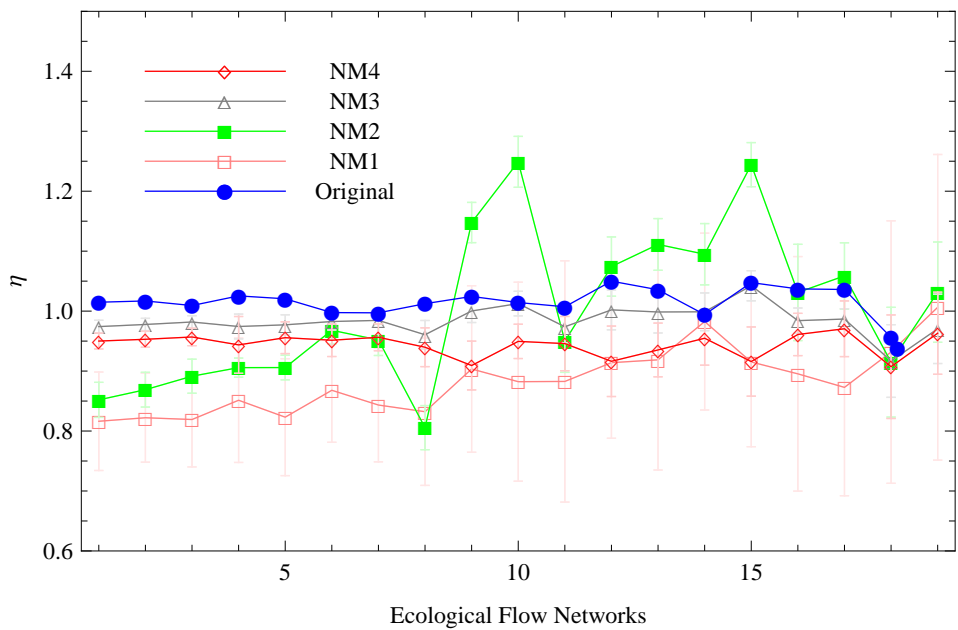}} \vskip3mm
\caption{Comparison for Allometric Scaling Exponents of Original
Networks with Null Models}{\small{All the empirical flow networks
are sorted in the order of Table \ref{tab.foodwebdata} along the
horizontal axis(the left most network has largest number of nodes).
The data points and error bars in all null models stand for the
average numbers and standard deviations of exponents $\eta$ of 50
experiments.}}\label{fig.nullmodelexponent}
\end{figure}

Furthermore, we generate 50 networks for each null model on each
empirical flow network collected. The average values of $\eta$s are
compared to the original ones in Figure \ref{fig.nullmodelexponent}.
From this figure, we can see that the exponents of NM3 and NM4 are
more close to the original networks with less fluctuations. That
means the weights information is more important than the structures
and the weight correlation play a minor role on allometric
exponents. Although NM3 and NM4 have similar exponents as the
original networks, all their values are smaller than the ones of the
original networks.

\begin{figure}
\centerline {\includegraphics{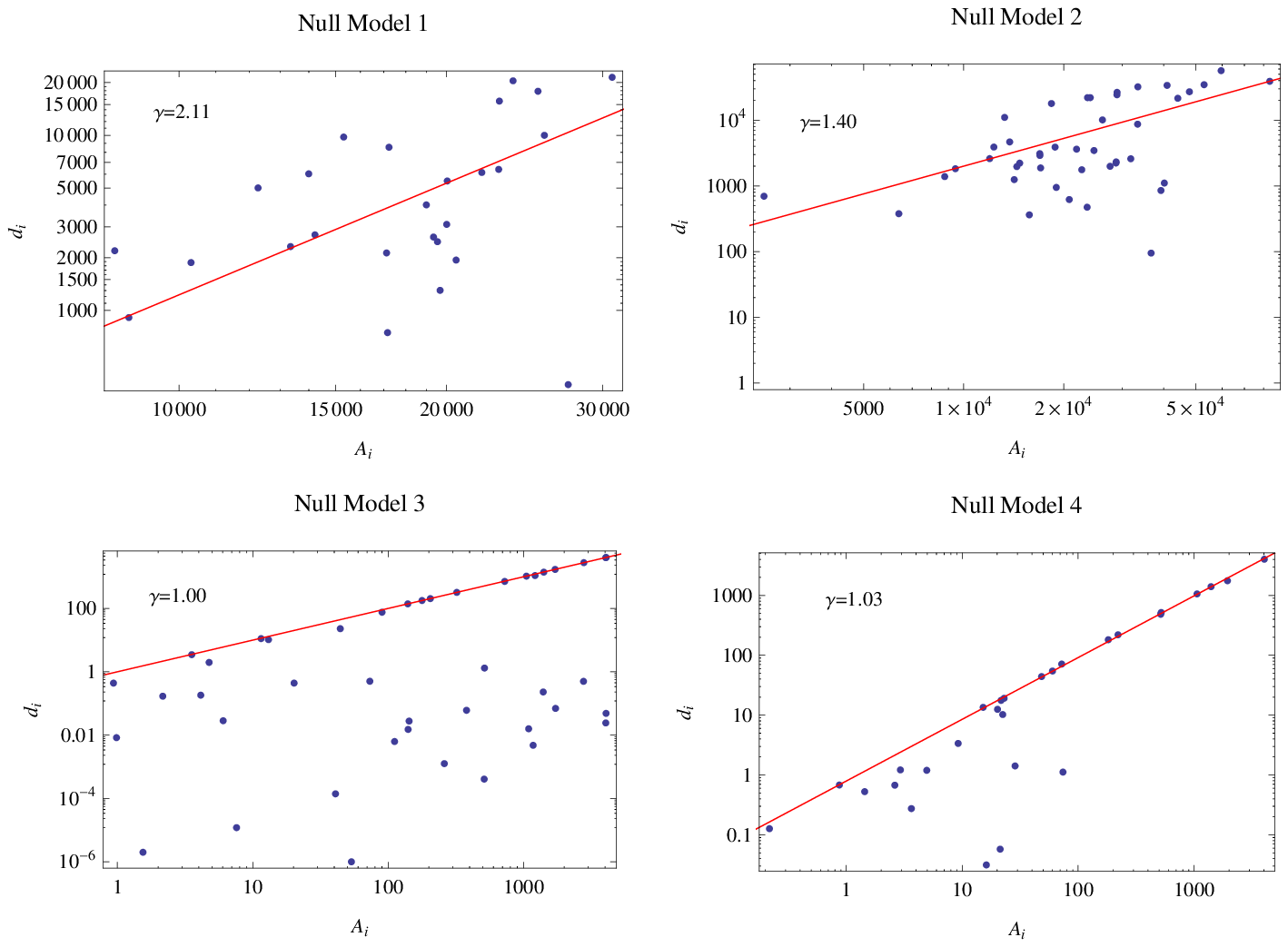}} \vskip3mm
\caption{Dissipation Law for Null Models of Mondego
Network}{\small{The dissipation laws are calculated for the
artificial balanced networks to compare with the allometric laws in
Figure} \ref{fig.nullmodels}}\label{fig.nullmodelgammas}
\end{figure}

Further studies on the dissipation laws of these null models can
explain the patterns shown in Figure \ref{fig.nullmodelexponent}.
From Figure \ref{fig.nullmodelgammas}, we observe that most networks
do not have obvious dissipation scaling law. However, we can observe
the obvious straight lines formed on the ceiling of data clouds for
NM3 and NM4, although lots of scatter points are below them.
Actually, the artificial balancing method can account for this
phenomenon. Because in NM3 and NM4, all the weights of links are not
changed but the connections are destroyed so that the energy influx
cannot balance with out flows for lots of nodes, the balanced flows
(dissipations) are almost proportional to the original flow.
Therefore, before we fit the data clouds by using a line on NM3 and
NM4, we actually left the data under the lines as outliers. In this
way, we can estimate the right $\gamma$s for NM3 and NM4 which are
close to 1. By comparing Figure \ref{fig.nullmodelgammas} and Figure
\ref{fig.nullmodels}, we know that the relationship between $\gamma$
and $\eta$ is also suitable for these null models.

\section{Flow Adjusting Algorithm}
\label{sec.flowadjustingalgorithm} In the main text Section
\ref{sec.gammaandeta}, we apply the so called ``Flow Adjusting
Algorithm'' to perturb the original flow network. In this section,
we will introduce the detailed steps of this method.

The main purpose is to conform the adjusted flow network to satisfy
two conditions (1) dissipation law with given exponent, and (2) flow
balance requirement without changing the topology and relative
weight of each flow. Or we can express it as a mathematical problem:
to find a solution of the following equations system.

\begin{equation} \label{eqn.eqnsystem}
\left\{ \begin{aligned}
         D_i=c (\sum_{j=0}^{N}{f_{ji}})^{\gamma} \\
                 \sum_{j=0}^{N}{f_{ji}}=\sum_{j=1}^{N}{f_{ij}}+D_i
                          \end{aligned} \right. \forall i \in [1,N]
\end{equation}

Where $c$ and $\gamma$ are given constants, $f_{ij}$s and $D_i$s are
variables.Therefore we have totally $2N$ equations but $E+N$
variables. For normal flow networks, $E>N$, so we should have
infinite solutions to the Equation \ref{eqn.eqnsystem}. However,
solving these equations are hard because they are non-linear.

The flow adjusting algorithm is an approximate algorithm to solve
these equations. At first, we have a nodes set $O$, a set of stop
criterions. Initially, we set time step $t=1$,
$O^{(1)}=\{1,2,\cdot\cdot\cdot,N\}$, the flux matrix as the original
flow network $F^{(1)}=F$, where the superscripts on $O$ and $F$ are
the current time step. The algorithm will repeat the following
steps:

(1). For any node $i$ in $O^{(t)}$, the algorithm needs the current
out flows from $i$, $\{f^{(t)}_{ij}|f^{t}_{ij}>0\}$. Solve the
equation for $x_i$:
\begin{equation}
x_i=\sum_{j=1}^{N}{f^{(t)}_{ij}}+c x_i^{\gamma},
\end{equation}
i.e., the new total influx to $i$.

(2). Assign $x_i$ to all incoming edges to $i$ proportionately, set
\begin{equation}
f^{(t+1)}_{ji}=x_i \frac{f^{(t)}_{ji}}{\sum_{j}{f^{(t)}_{ji}}},
\forall j\in \{j|f^{(t)}_{ji}>0\}
\end{equation}

and

\begin{equation}
D^{(t+1)}_{i}=(\frac{x_i}{c})^{\gamma}.
\end{equation}

(3). Add all input nodes $j$s of $i$ into set $O^{(t+1)}$ and delete
$i$ from $O^{(t+1)}$;

(4). Set $t=t+1$, repeat the previous steps until the stop
criterions, which include the total running time being smaller than
a given number, the dissipation law exponent $\gamma$ and $R^2$ for
the new flux matrix $F^{(t)}$ being close to the wanted values, are
satisfied.

This algorithm works once the network is connected (which means
there is at least one path from $0$ to every node $i$). For most of
networks, the algorithm can converge to a network pertaining the
significant dissipation law on the given $\gamma$ exponent. However,
it may oscillate on some topologies, especially for random networks.
Further approvement of this algorithm will leave for the future
works.

\section{$\gamma$ and $\eta$ Relationships for Modeled Networks}
\label{sec.simulatednetworks} To better understand how $\gamma$
correlates with $\eta$, we will study several special modeled
networks in this section.
\subsection{Minimum Spanning Tree}
\label{sec.mst} We now consider a special case: minimum spanning
tree introduced by \cite{Frank_simple_2005}. By controlling two
parameters $\theta$ and $\beta$, we can generate variant trees with
different basal species ratio(controlled by $\beta$) and maximum
trophic level (controlled by $\theta$).

The tree's construction process is as follows. Let's consider an
ecological community with $S$ different species, in which a
hypothetic food web (tree structure) will be built. At first, we
select $\beta S$ species as the basal species at the first trophic
level. And at each time, a new species $j$ is added to the minimum
spaning tree. $j$ will select a node $i$ as its unique prey
according to the probability:
\begin{equation}
P_{ij}=\frac{t_j^{-\theta}}{\sum_{k\in T}{t_k^{-\theta}}},
\end{equation}
where, $t_j$ is $j$'s trophic level + 1 (i.e., the depth of $j$ in
the tree), $T$ is the set of species which are already in the
spanning tree, and $\theta$ is a parameter to control the attachment
preference of the new node on depth. If $\theta$ is large, the new
node may attache to the position being close to the root.

After a tree is constructed, we will assign random values in the
original flux matrix and then apply the ``Flow Adjusting Method'' on
it. In this way, we can investigate the influence of both tree's
structure ($\theta$ and $\beta$) and dissipation exponent $\gamma$
on the allometric scaling exponent $\eta$.

\begin{figure}
\centerline {\includegraphics[scale=0.6]{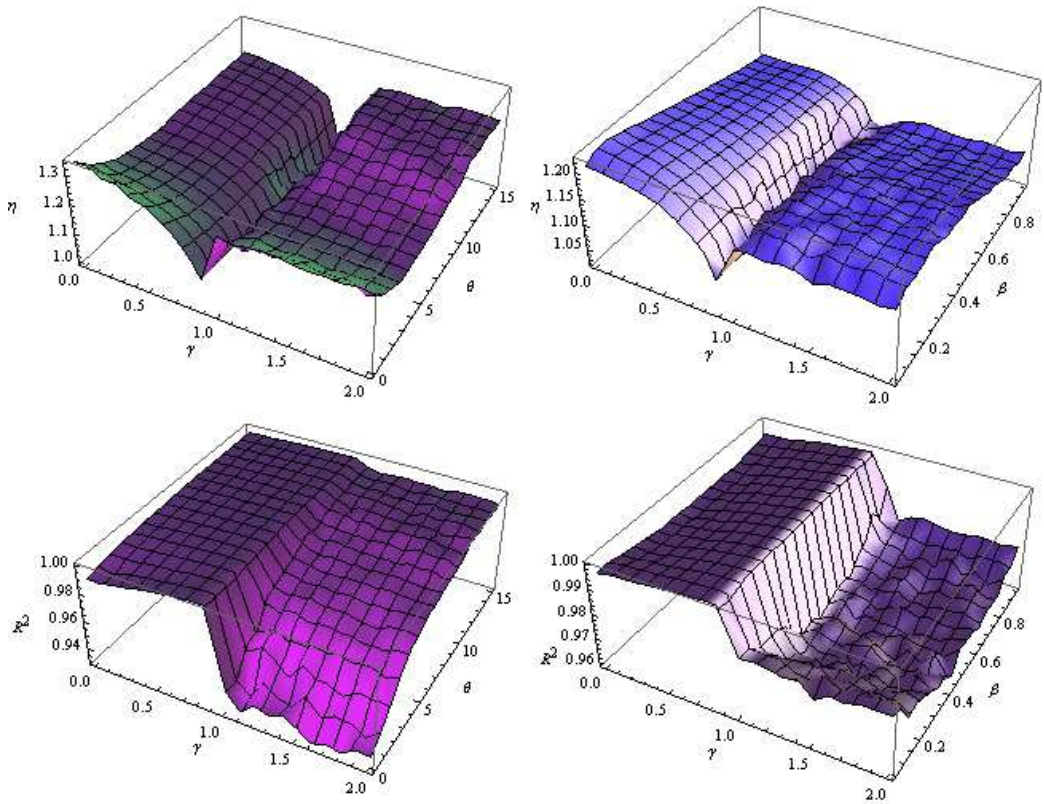}} \vskip3mm
\caption{$\eta$ and $R^2$ Change with $\theta$,$\beta$ and
$\gamma$}{\small{For each combination of parameters, we generate 10
minimum spanning trees to get the average value of $\eta$. The
species number $S=100$ in all simulations}}\label{fig.3dtree}
\end{figure}

From Figure \ref{fig.3dtree}, we found at first both the dissipation
law and network structure can affect the allometric scaling.
However, $\eta$ depends more on $\gamma$ than $\beta$ and $\theta$
because $\eta$ will change with $\gamma$ intensively. When $\gamma$
is given, we can observe the similar trend of $\eta$ depends on
$\beta$ and $\theta$ as the results introduced by
\cite{Frank_simple_2005}.

Interestingly, when $\gamma$ is set to be 0, each node's dissipation
is a constant, this corresponds to Garalaschelli's approach's
assumption (see Figure \ref{fig.hfoodweb}(d)). And the exponent
$\eta$ derived by our algorithm is exactly same as the result
derived by Garlaschelli's approach on the same tree. And the
dependence of $\eta$ on $\beta$ and $\theta$ is same as
\citep{Frank_simple_2005}. So, our method can recover Garlaschelli's
method on spanning trees.

\subsection{Random Network}
\begin{figure}
\centerline {\includegraphics[scale=1.1]{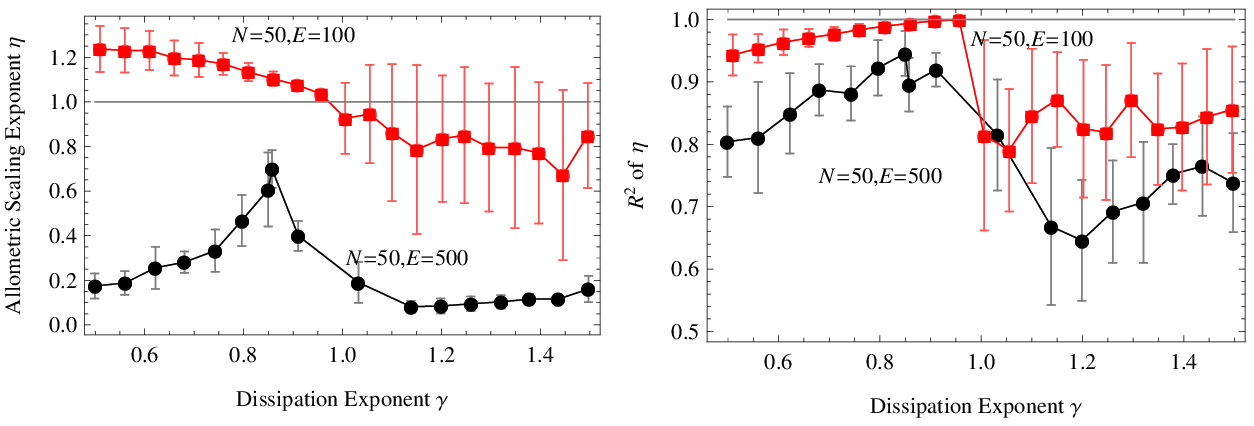}}
\vskip3mm \caption{$\eta$ and $R^2$ Change with $\gamma$ for
different random networks}{\small{All $\eta$s are averaging for 10
random networks}}\label{fig.random}
\end{figure}

We also test the ``Flow Adjusting Algorithm'' on random networks.
The results are shown in Figure \ref{fig.random}.

We only show the results of random networks with 50 nodes but
different number of edges because the FAA is hardly to converge on
the random networks with large number of edges. When the algorithm
cannot get a final result after a given number of time steps (200),
we have to regenerate a new random network with the same number of
nodes and edges. It is interesting to observe that when the number
of edges is large, the responding curve of $\eta$ on $\gamma$ is
very different from the ones in Figure \ref{fig.gammaandeta}.
$\eta$s are always very small, where $\eta$ almost gets a peak when
$\gamma$ approaches $1$. Therefore, we know the topological
structure does affect the allometric exponent.

Furthermore, we generate random networks based on a minimum spanning
tree by adding $\alpha (N(N-1)/2-N)$ additional edges randomly. When
$\alpha=0$, the network is an MST, while when $\alpha=1$, it is a
complete graph. Therefore, we can observe how the corresponding
curve of $\eta$ on $\gamma$ changes when the network structure
changes from a tree to a random network by tuning $\alpha$.
\begin{figure}
\centerline {\includegraphics[scale=0.6]{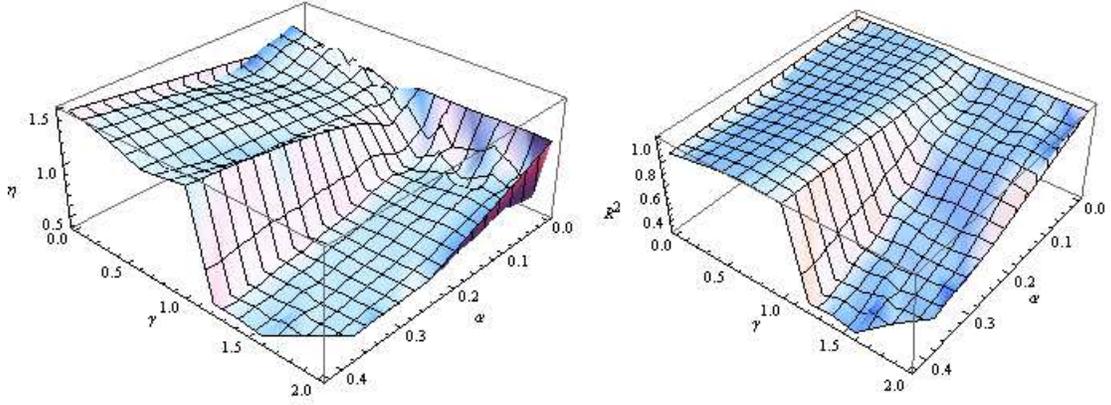}} \vskip3mm
\caption{$\eta$ and $R^2$ Change with $\gamma$ and $\alpha$ on the
Random Networks based on Spanning Trees}{\small{All $\eta$s are
average results of 10 networks, the backbone trees are of the
parameters $N=100,\beta=0.5,\theta=5$.}}\label{fig.tree2rand}
\end{figure}

From Figure \ref{fig.tree2rand}, the dependance of $\eta$ on
$\gamma$ when $\alpha=0.4$ is different from the random networks
with the same connectance but similar to the ones of empirical
ecological networks. That is because the former random network is
generated based on a minimum spanning tree which can be viewed as
its backbone. Therefore, the spanning tree backbone is a key
ingredient to the allometric exponent. Additionally, in Figure
\ref{fig.tree2rand}, all the random networks based on spanning tree
with different $\alpha$ have a peak on $\eta$ when $\gamma=1$.

According to these experiments, we know both network structure and
dissipation law exponent can influence allometric exponents. But the
exponent $\gamma$ is more important than the structure. And the
shape of backbone spanning trees can change the shape of the
relationship between $\eta$ and $\gamma$.

\section{Flow Adjustment for Empirical Ecological Flow Networks}
\label{sec.empiricalnetwork} ``Flow Adjusting Algorithm'' is applied
to collected empirical ecological flow networks as shown in Figure
\ref{fig.gammaandeta} on Mondego as an example. In this section, we
will show the results for other networks and discuss the technique
details.

\begin{figure}
\centerline {\includegraphics[scale=1]{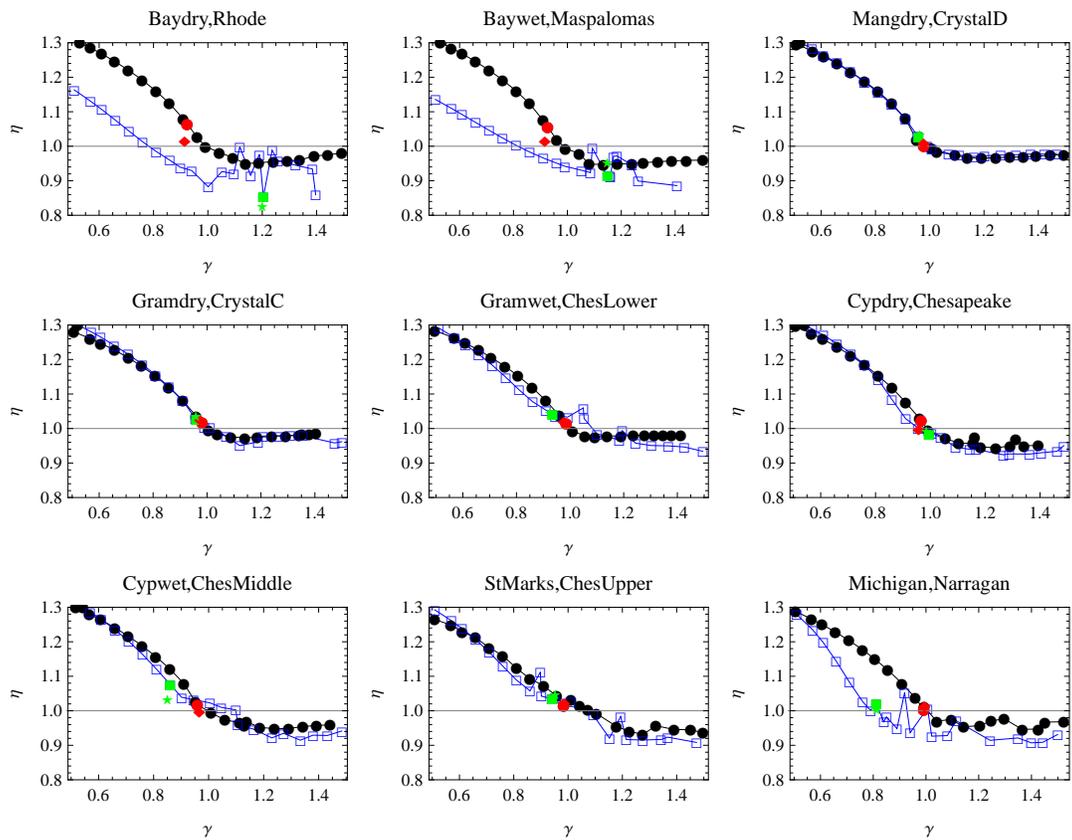}}
\vskip3mm \caption{$\gamma$ and $\eta$ Relation Adjusted by ``Flow
Adjusting Algorithm'' on Collected Ecological Networks}{\small{Each
plot shows results of ``Flow Adjusting Algorithm'' on two empirical
networks (black and blue curves), the red diamond and disk stand for
the $\gamma$,$\eta$ combinations for the original exponents and the
adjusted results by ``Flow Adjusting Algorithm'' on the original
$\gamma$ for the first ecological network; and the green star and
square are for the second ecological
network.}}\label{fig.allfoodwebsgammaeta}
\end{figure}

The basic idea of ``Flow Adjusting Algorithm'' is to tune various
flows on the network until the dissipation law exponent is close to
the wanted value. When we apply this method on the empirical
ecological networks, we adjust the flows until either (1) the
adjusted dissipation law exponent $\gamma'$ satisfies
$|\gamma'-\gamma^*|<0.01$, where $\gamma^*$ is the wanted exponent
and (2) the $R'^2$ of the dissipation power law on the adjusted
flows should satisfy $R'^2\leq R_{\gamma}^2$, where $R_{\gamma}^2$
is the Rsquare of the dissipation law for the original flow network;
or (3) The running time steps are larger than 500. Because the
algorithm may diverge, whence the first requirement may not be
satisfied, we have to stop the algorithm within a finite time steps
and retrieve one of the best network as the output.

In Figure \ref{fig.allfoodwebsgammaeta}, we show results for
applying FAA on all collected ecological networks. The red diamond
and green stars are the $\gamma$ and $\eta$ combinations for the
original networks, and the red disk and green squares stand for the
ones for FAA results on the original $\gamma$ as the designed
exponent. If the original flow network satisfies the dissipation law
and flow balance condition perfectly, then the adjusted $\eta$ value
should be similar with the value of the original network (which
indicates that the red disks(green squares) should overlap with the
red diamonds(green stars)). However, we observe that the markers for
some networks do not overlap which means the corresponding original
networks do not satisfy that two conditions perfectly.

\begin{figure}
\centerline {\includegraphics[scale=1]{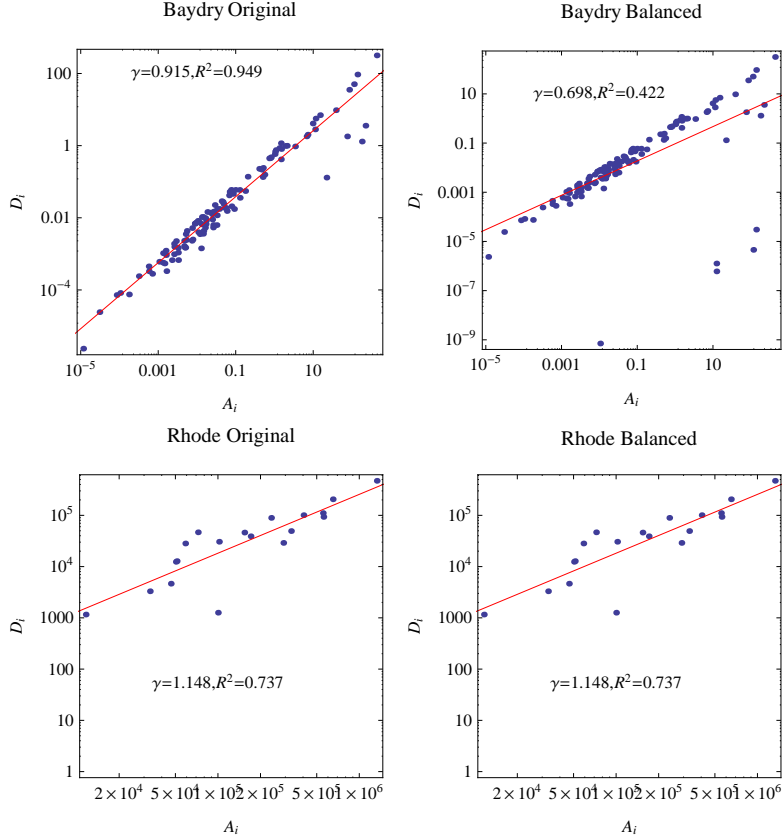}} \vskip3mm
\caption{Dissipation Scaling Law for Original and Balanced Flow
Networks of Baydry and Rhode}\label{fig.baydryrhode}
\end{figure}
Let's see Baydry network as an example. Although the original flow
network possess a very good dissipation law as shown in the left-top
plot in Figure \ref{fig.baydryrhode}, this network is not balanced.
Because balance is a basic requirement of the algorithm for
allometric scaling law mentioned in Section
\ref{sec.allometricscaling}, we have to balance Baydry network at
first by the method in Section \ref{sec.balance}. However, the
balanced flow network always has different dissipation law (compare
the left-top plot to the right-top one in Figure
\ref{fig.baydryrhode}). Therefore, the original $\eta$ and the one
adjusted by FAA are different (The first plot red disk and diamond
in Figure \ref{fig.allfoodwebsgammaeta}).

Another example network which is balanced but not be of good
dissipation law is Rhode as shown in the left-bottom and
right-bottom plots of Figure \ref{fig.baydryrhode}. We observe that
the dissipation scaling law is not significant ($R^2=0.737$) for
Rhode network. Therefore, the original and adjusted exponents do not
overlap (The first plot green star and square in Figure
\ref{fig.allfoodwebsgammaeta}). Furthermore, FAA cannot obtain a
convergent result on Rhode network, that is the reason why the green
star and square have different horizontal coordinates.

Because the negative relationship between $\gamma$ and $\eta$ is
significant only if the flow network satisfies (1) a significant
dissipation law and (2) the flows are balanced, we adjust the flows
of the empirical networks to conform the dissipation law with the
original dissipation law. In this way, we believe the noise
contained in original data can be eliminated, so that more exact
allometric exponents can be computed. The red solid curve in Figure
\ref{fig.gammaandeta} shows the original $\gamma$ exponents and
adjusted $\eta$ exponents.

%
%
%
%

\end{document}